\documentclass[conference]{IEEEtran}
\usepackage{cite}
\usepackage{url}
\usepackage{amsmath,amssymb,amsfonts}
\usepackage{algorithmic}
\usepackage{graphicx}
\usepackage{lipsum}
\usepackage{hhline}
\usepackage{tabularx}
\usepackage{xcolor}
\usepackage{multirow}
\usepackage{subcaption}
\usepackage{dblfloatfix}    
\usepackage[font=small]{caption}

\graphicspath{{graphics/}}
\usepackage{algorithm}
\usepackage{algorithmic}
\usepackage{verbatim}
\usepackage{cprotect}

\makeatletter
\def\url@leostyle{%
	\@ifundefined{selectfont}{\def\UrlFont{\sf}}{\def\UrlFont{\small\ttfamily}}}
\makeatother
\DeclareMathOperator*{\argmin}{arg\,min} 

\def\BibTeX{{\rm B\kern-.05em{\sc i\kern-.025em b}\kern-.08em
		T\kern-.1667em\lower.7ex\hbox{E}\kern-.125emX}}
\begin{document}
	
	\title{Attention Aided CSI Wireless Localization \\
		\thanks{}
	}
	
	\author{\IEEEauthorblockN{1\textsuperscript{st} Given Name Surname}
		\IEEEauthorblockA{\textit{dept. name of organization (of Aff.)} \\
			\textit{name of organization (of Aff.)}\\
			City, Country \\
			email address or ORCID}
		\and
		\IEEEauthorblockN{2\textsuperscript{nd} Given Name Surname}
		\IEEEauthorblockA{\textit{dept. name of organization (of Aff.)} \\
			\textit{name of organization (of Aff.)}\\
			City, Country \\
			email address or ORCID}
		\and
		\IEEEauthorblockN{3\textsuperscript{rd} Given Name Surname}
		\IEEEauthorblockA{\textit{dept. name of organization (of Aff.)} \\
			\textit{name of organization (of Aff.)}\\
			City, Country \\
			email address or ORCID}
	}
	
	\author{
		\IEEEauthorblockN{Artan Salihu$^\dagger$$^\ddagger$, Stefan Schwarz$^\dagger$$^\ddagger$ and Markus Rupp$^\dagger$}
		\IEEEauthorblockA{$^\dagger$ Institute of Telecommunications, Technische Universit{\"a}t (TU) Wien\\
			$^\ddagger$ Christian Doppler Laboratory for Dependable Wireless Connectivity for the Society in Motion \\
			Email: \{artan.salihu,stefan.schwarz,markus.rupp\}@tuwien.ac.at\\
		}
	}
	
	\maketitle
	\begin{abstract}
		Deep neural networks (DNNs) have become a popular approach for wireless localization based on channel state information (CSI). A common practice is to use the raw CSI in the input and allow the network to learn relevant channel representations for mapping to location information. However, various works show that raw CSI can be very sensitive to system impairments and small changes in the environment. On the contrary, hand-designing features may hinder the limits of channel representation learning of the DNN. In this work, we propose attention-based CSI for robust feature learning. We evaluate the performance of attended features in centralized and distributed massive MIMO systems for ray-tracing channels in two non-stationary railway track environments. By comparison to a base DNN, our approach provides exceptional performance.\vspace{-0.25cm}
	\end{abstract}
	
	\vspace{-0.2cm}
	\section{Introduction}
	The deployment of massive multiple-input multiple-output (MIMO) technology in the fifth generation (5G) mobile cellular systems can enable high-accuracy positioning services, where ambitious meter-level accuracy requirements are set \cite{3gpp.22.261}. Recently, deep learning has become a renowned approach for achieving exceptionell localization performance \cite{9348191, 9616218, sun2019fingerprint}. Such DNN-based methods take advantage of a large amount of channel state information (CSI) available at a massive MIMO base station (BS) to train a model with the channel prints from the known locations. The model then utilizes the channel estimates of the unknown transmitter to determine its position related information. 
	
	A common approach is to use the raw CSI as an input to the DNN architecture. However, the raw CSI can be very sensitive to system impairments and slight variations in the environment. Thus, we might require a vast number of location-tagged CSI to achieve sufficiently rich representation learning of distinct locations. A variety of works have addressed the issue of imperfect channel estimates by suggesting to hand-design more robust features, mainly by exploiting the approximately sparse angle- and delay-domain channel representation in a MIMO-OFDM system. For instance, the work in \cite{9348191} suggests a decimated delay-domain CSI representation followed by autocorrelation to capture features that are invariant to the system impairments. Similarly, the work in \cite{sun2019fingerprint} suggests utilizing angle-delay channel representation as input to a convolutional neural network (CNN) based model. However, hand designing the input features hinders the limits of the DNNs for achievable representation learning of the channel.
	
	Alternatively, we can improve the feature learning process at the beginning of the DNN itself by leveraging the attention mechanism \cite{graves2013generating} and allowing the neural network to \textit{attend} on different parts of the input. The attention module is at the core of every Transformer architecture. The Transformer was initially proposed in \cite{vaswani2017attention} for natural language processing (NLP) and recently has been successfully applied as an alternative to CNNs in computer vision \cite{dosovitskiy2020image}. While the attention mechanism has become a \textit{de facto }standard for signal processing in NLP and vision, its ability for CSI feature learning in wireless communications and wireless localization, in particular, remains underexplored.\vspace{-0.2cm}
	\begin{figure}[!t] 
		\centering
		{%
			\includegraphics[width=0.82\linewidth]{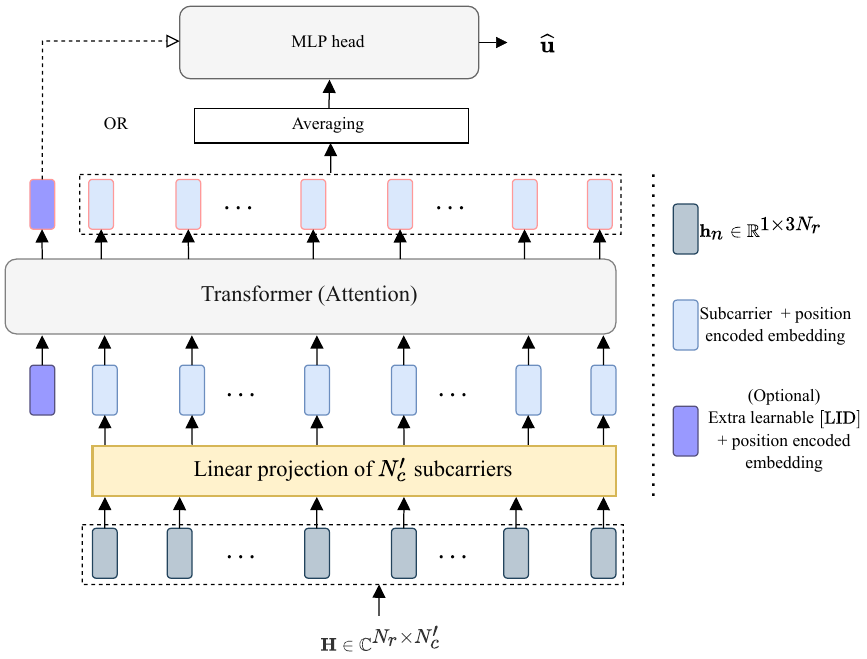}}	
		\cprotect\caption{Overview of the attention-aided model. We linearly embed each subcarrier, add position embeddings, and feed the representation vectors to a Transformer-like block with an attention module for feature extraction. For location estimation, we average over the attended features. Instead, we can use an extra learnable $\left[ \verb*|LID| \right] $ embedding too.} \vspace{-0.3cm}
		\label{fig:WiT_architecture_overview} \vspace{-0.4cm}
	\end{figure}
	\subsection*{Our Contributions}
	In this work, we firstly propose an efficient and a general robust feature learning process incorporated into an end-to-end DNN. Our model is based on the attention mechanism, which serves as an adaptive filter for CSI features resilient to imperfect channel estimates and temporal variations in the environment. To achieve both, robustness and scalability, we show that we can use a Transformer-like architecture to feed the whole channel estimate without using convolution layers, fusion approaches, recurrence, or decimating the input channel. An overview of the proposed model is depicted in Fig. \ref{fig:WiT_architecture_overview}.
	Secondly, we provide a comprehensive evaluation of the proposed method by applying it to ray-tracing channels obtained along two railway tracks in carefully modeled changing surrounding environments. Finally, we present insights regarding localization accuracy in a centralized and distributed antenna system. \vspace{-0.4cm}	
	\section{System Model}\label{systemModel}
	We consider uplink transmission over orthogonal frequency-division multiplexing (OFDM) in a massive MIMO system. We assume $R$ single-antenna transmitters placed in a space $\mathbb{R}^{3}$ at positions $\mathbf{u}_{r} = [u_{r,1},u_{r,2},u_{r,3}]^{T}$ with $r \in \mathcal{R}$, where $\mathcal{R}$ denotes the set of user location indices and $\left| \mathcal{R} \right| = R$. The base station (BS) is equipped with $N_{r}$ antenna elements. Alternatively, we also consider $N_{r}$ spatially distributed antennas among $M$ remote radio heads (RRHs) at positions $\mathbf{b}_{m} = [b_{m,1},b_{m,2},b_{m,3}]^{T} $ with $m\in \mathcal{M}$, where $\mathcal{M}$ is the set of RRH indices and $\left|\mathcal{M}\right| = M$. In case of distributed antennas, we assume that all RRHs are connected via high-speed fronthaul links to the central unit (CU), i.e., the delay between the RRHs and the CU is negligible. Further, we consider $S$ scattering objects in the ROI at respective positions, $\mathbf{p}_{s} = [p_{s,1},p_{s,2},p_{s,3}]^{T}$ with $s \in \mathcal{S}$, where $\mathcal{S}$ denotes the set of scattering object indices and $\left| \mathcal{S} \right| = S$. \vspace{-0.0cm}
	
	\subsection{Dynamic Environment}
\begin{figure}[!h] 
		\centering
		\subfloat[An example of $\mathrm{RMS}$ delay spread. \label{fig:rms_delay}]{%
			\includegraphics[width=0.96\linewidth]{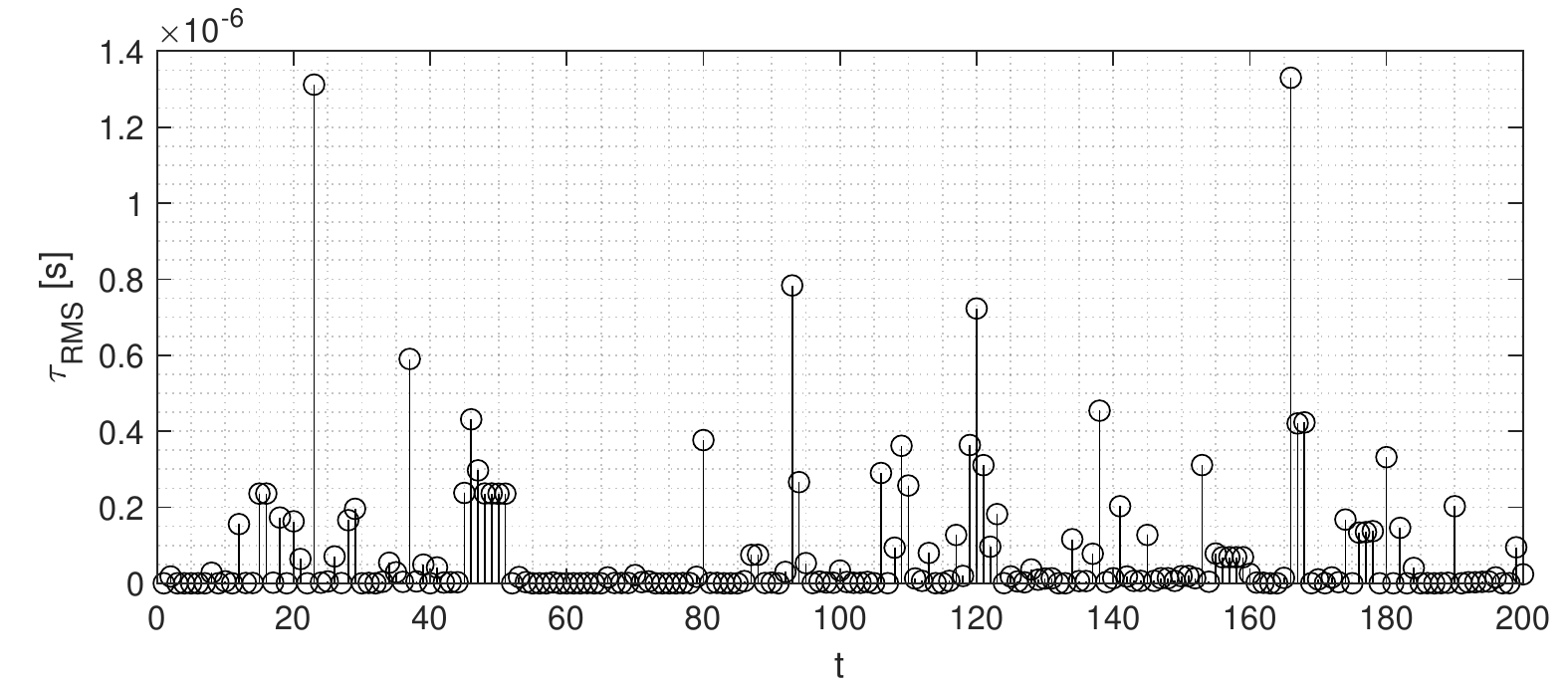}}
		\hfill
		\\
		\vspace{-0.0cm}
		\subfloat[An example of $\mathrm{RMS}$ angle of arrival spread in azimuth \label{fig:rms_angle}]{%
			\includegraphics[width=0.96\linewidth]{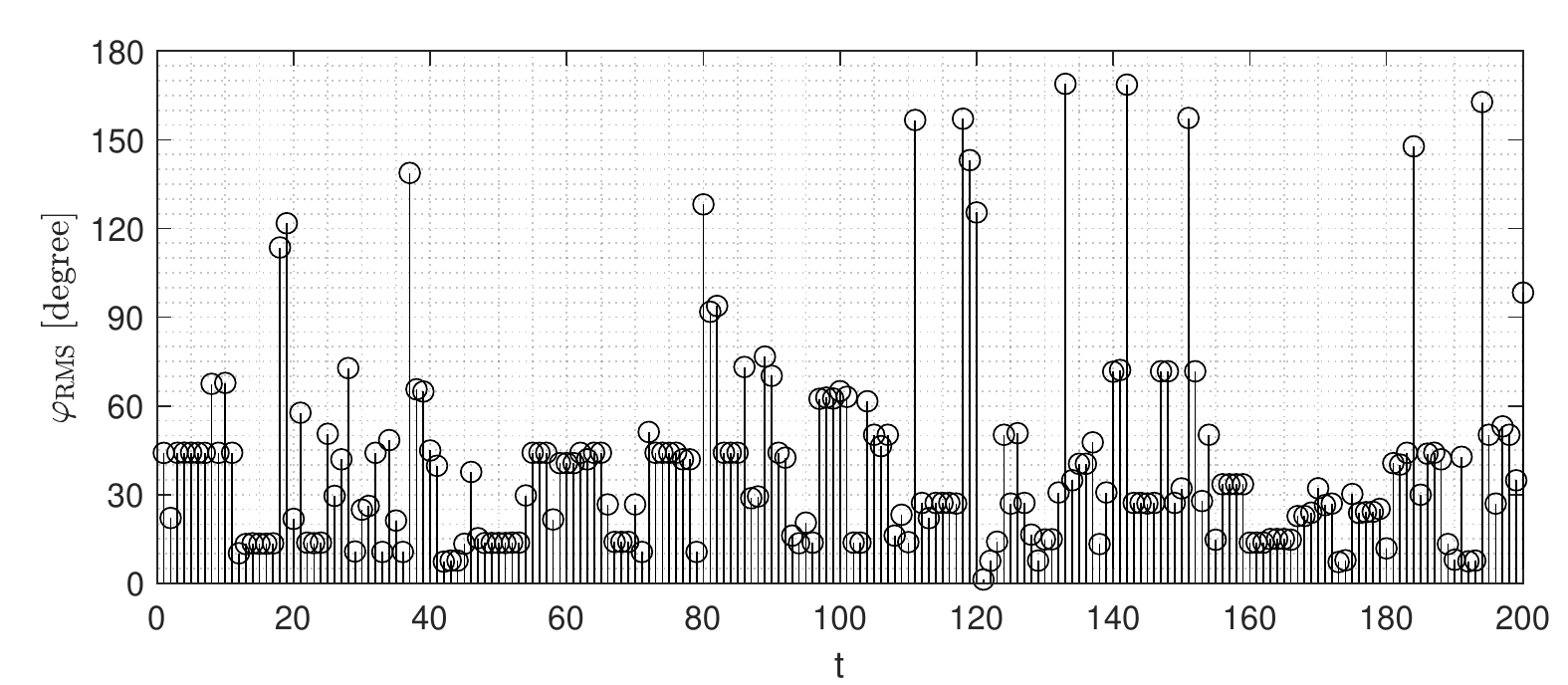}}
		\vspace{0.0cm}\caption{Delay- and angle-spread at $\mathbf{u}_{r}$ over $T=200$ time snapshots.}
		\label{fig:rms_dynamic_scenario}\vspace{-0.2cm}
	\end{figure}

	In this paper, we consider that the propagation environment changes in each time snapshot $t \in \{1, \ldots, T\}$. More specifically, by fixing the positions of the receiver and transmitter, we realize the time-varying conditions of the environment by altering the positions of $S'$ scatterering objects, where $ S' = \left| \mathcal{S'} \right|$ and $ \mathcal{S'} \subseteq \mathcal{S}$. Thus, we have 	\vspace{-0.1cm}
	\begin{equation}
		{p}_{s,i}^{t} = {p}_{s,i} + {z}_{s,i},
	\end{equation} 
	where ${z}_{s,i}$ is the zero-mean Gaussian noise with variance $\mathbf{\sigma}_{z}^{2}$ at $i-$th coordinate. Similarly, we account for the uncertainty in the position of antenna of the transmitter at $t$, $\mathbf{u}_{r}^{t}$, i.e., \vspace{-0.2cm}
	\begin{equation}
		{u}_{r,i}^{t} = {u}_{r,i} + {n}_{r,i},
	\end{equation} 
	where ${n}_{r,i}$ is the zero-mean Gaussian noise with variance ${\sigma}_{n}^{2}$ at $i-$th coordinate. Note that the variations in the position of the scatterers alter the gain, delay and angle information of the individual multi-bounce non-line-of-sight (NLOS) paths. In Fig. \ref{fig:rms_dynamic_scenario}, we show an example of the $\mathrm{RMS}$ delay-spread,  $\tau_{\text{RMS}}$, as well as the $\mathrm{RMS}$ angle of arrival spread in azimuth,  $\varphi_{\text{RMS}}$, for a single random $\mathbf{u}_{r}$ over $T = 200$ time snapshots and $L = 4$ strongest paths. Here, the delay is normalized with respect to the strongest path. Moreover, the uncertainty in the position of antenna, allows us to account for the effect of imperfect channel estimates at the receiver. In Sec. \ref{channel_model}, we detail the geometric channel model and the relationship with the position parameters, where recognizing the impact of the antenna position offset in the channel is easy to perceive.
	
	Additionally, to keep this work more general, we consider that the electromagnetic properties of the scattering objects change over time, which impacts the amplitude gain of the radar cross section (RCS) of the scattering objects. We assume that material types change randomly and have a permittivity value of $\epsilon_{\kappa}$ at time $t$ with $\kappa \in \mathcal{K}$, where $\mathcal{K}$ is the set of material types. Finally, we also consider atmospheric attenuation in the environment. Thus, in case of a rainy period $\boldsymbol{\mathcal{R}}$ with the probability $\mathbb{P}  \left( \boldsymbol{\mathcal{R}} \right)$, we assume additional attenuation to the line-of-sight (LOS) path.\vspace{0.1cm}  
	
	\subsection{Channel Model}\label{channel_model}
	We assume that the signal from each transmitter $r$ is received at $N_{r}$ antennas over a set of active subcarriers $\mathcal{N}_{c}' = \{ k_{1}, \ldots, k_{N_{c}'} \}$ subcarriers. The frequency-domain channel for the $k-$th subcarrier reads \cite{heath2016overview}
	\begin{equation}
		\widehat{\mathbf{h}}_{k}^{t} = \sum_{\ell=1}^{L} \eta_{\ell}^{t} e^{j{2 \pi k}{\Delta f} \tau_{\ell}^{t}} \mathbf{a}\left(\varphi_{a z,\ell}^{t}, \varphi_{e l,\ell}^{t} \right) . \vspace{-0.05cm}
	\end{equation}
	In above, $\eta_{\ell}^{t}$ and $\tau_{\ell}^{t}$ denote the $\ell-$th path's complex gain and propagation delay at time $t$. The subcarrier spacing is $\Delta f = B/N_{c}$, where $B$ is the bandwidth and $N_{c}$ is the total number of subcarriers. The angles of arrival (AoA) in azimuth and elevation are denoted by $\varphi_{a z,\ell}^{t}$ and $\varphi_{e l,\ell}^{t}$, respectively. The expression for the steering vector at the receiver is given by 
	\begin{equation}
		\mathbf{a}\left(\varphi_{\mathrm{az}}^{t}, \varphi_{\mathrm{el}}^{t}\right)=\mathbf{a}_{z}\left(\varphi_{\mathrm{el}}^{t}\right) \otimes \mathbf{a}_{x}\left(\varphi_{\mathrm{az}}^{t}, \varphi_{\mathrm{el}}^{t}\right) \vspace{-0.05cm}.
	\end{equation}
	The array steering vectors $\mathbf{a}_{x}(\cdot)$, and $\mathbf{a}_{z}(\cdot)$ are \vspace{-0.0cm}
	\begin{equation}
		\begin{aligned}
			\mathbf{a}_{x}\left(\varphi_{\mathrm{az}}^{t}, \varphi_{\mathrm{el}}^{t}\right)=\left[1, e^{j \frac{2\pi}{\lambda_c} d \sin \left(\varphi_{\mathrm{el}}^{t}\right) \sin \left(\varphi_{\mathrm{az}}^{t}\right)}, \ldots\right.\\ 
			\left.\ldots, e^{j \frac{2\pi}{\lambda_c} d\left(M_{x}-1\right) \sin \left(\varphi_{\mathrm{el}}^{t}\right) \sin \left(\varphi_{\mathrm{az}}^{t}\right)}\right]^{T},
		\end{aligned}
	\end{equation}
	\vspace{-0.4cm}
	\begin{gather*}
		\mathbf{a}_{z}\left(\varphi_{\mathrm{el}}^{t}\right)=\left[1, e^{j \frac{2\pi}{\lambda_c} d \cos \left(\varphi_{\mathrm{el}}^{t}\right)}, \ldots, e^{j \frac{2\pi}{\lambda_c} d\left(M_{z}-1\right) \cos \left(\varphi_{\mathrm{el}}^{t}\right)}\right]^{T}, \vspace{-0.0cm}
	\end{gather*}
	with $\lambda_{c} = c/f_{c}$, where $f_{c}$ being the carrier frequency and $c$ the speed of light, and $d = \lambda_{c}/2$ the antenna element spacing. In this work, we use a ray-tracer to obtain the delays, angles and path gains.
	The channel for the $r-$th user location $\mathbf{H}_{r}^{t}$ over $N_{c}'$ subcarriers can then be written as
	\begin{equation}
		\mathbf{H}_{r}^{t} =\left[\mathbf{\widehat{h}}_{k_1}^{t}, \mathbf{\widehat{h}}_{k_2}^{t}, \ldots, \mathbf{\widehat{h}}_{k_{N_{c}'}}^{t}\right] \in \mathbb{C}^{N_{r} \times N_{c}'} .
	\end{equation}
	For convenience, we drop the superscript $t$ from our notation.
	\section{Attention Aided Localization}\label{AttentionAided}
	\subsection{Problem Formulation}
	\vspace{0.1cm}
	Our goal in this work is to solve the problem of localization of the user from the obtained imperfect channel estimates in a changing surrounding environment. To do so, we rely on deep learning and formulate the DNN as a function $f^{\boldsymbol{\Psi}}(\cdot) $ parameterized by $\boldsymbol{\Psi}$ where, given the input channel $\mathbf{H}_{r}$, we aim to learn a set of robust features and directly map them into a position estimation, $\mathbf{\widehat{u}}_{r}$. The set of optimal parameter values $\boldsymbol{\Psi}$ is learned by minimizing a given loss function, \vspace{-0.0cm}
	\begin{equation}
		\begin{aligned}
			\argmin J(\mathbf{H}_r,\mathbf{u}_r;\boldsymbol{\Psi}); \quad  J = \mathbb{E}\left\|f^{\boldsymbol{\Psi}}(\mathbf{{H}}_{r})-\mathbf{u}_{r} \right\|^{2}.\vspace{-0.0cm}
		\end{aligned}\vspace{0.0cm}
	\end{equation}
	
	\subsection{Input Channel Representation}
	\vspace{0.1cm}
	We view the received channel matrix $\mathbf{H}_{r}$ as a set of $N_{c}'$ channel vectors of size $\mathbf{\widehat{h}}_{n}\in \mathbb{C}^{1 \times N_{r}}$, where $n \in \mathcal{N}_{c}'$. We handle the complex-valued CSI as three independent real numbers, i.e., $\mathbf{\widehat{h}}_{n}^{(\operatorname{Re})} = \operatorname{Re}\left\{\mathbf{\widehat{h}}_{n}\right\}$, $\mathbf{\widehat{h}}_{n}^{(\operatorname{Im})} = \operatorname{Im}\left\{\mathbf{\widehat{h}}_{n}\right\}$, and $\mathbf{\widehat{h}}_{n}^{(\operatorname{Abs})} = \operatorname{Abs}\left\{\mathbf{\widehat{h}}_{n}\right\}$, representing the real, imaginary and absolute parts. Additionally, the whole dataset, i.e., both the training and testing sets, is scaled by dividing each part with the maximum absolute value in it, $\Delta_{\operatorname{Re}} = \max(\max(\{\{|\mathbf{{H}}_{r}^{t,(\operatorname{Re})}|\}_{r=1}^{R}\}_{t=1}^{T}))$. Similarly, we normalize the imaginary, $\Delta_{\operatorname{Im}}$, as well as the absolute part, $\Delta_{\operatorname{Abs}}$. The input representation of each subcarrier for the network depicted in Fig. \ref{fig:WiT_architecture_overview} becomes $\mathbf{h}_{n} \in \mathbb{R}^{1 \times 3N_{r}}$.\vspace{0.2cm}
	
	\subsection{Transformer and Attention}
	\vspace{0.1cm}
	We maintain the same number of $D$ hidden units across the Transformer block shown in Fig. \ref{fig:attention_module}. Therefore, we firstly project each subcarrier into an embedding through a linear layer with learnable parameters $\mathbf{E} \in \mathbb{R}^{3N_{r} \times D}$, i.e.,\vspace{-0.1cm}
	\begin{equation}
		\begin{aligned}
			\mathbf{e}_{i} = \mathbf{h}_{i}\mathbf{E}. 
		\end{aligned}\vspace{-0.1cm}
	\end{equation}
	A characteristic of the attention module is that it is permutation-equivariant concerning the input embedded subcarriers. However, the structure of the whole channel, i.e., the arrangement, can reveal meaningful correlations among frequency-selective subcarriers. Thus, as we use no recurrence and no convolution, we inject some information about the indices of the subcarriers into the model.
	\subsubsection{Subcarrier Positional Encoding}
	We rely on absolute positional encoding \cite{gehring2017convolutional} to represent the arrangement of subcarriers. Specifically, we assign a learnable real-valued vector embedding $\mathbf{g}_{i}\in \mathbb{R}^{1\times D}$ to each subcarrier index $i$. Then, given the input channel, $\mathbf{g}_{i}$ is added to the subcarrier embedding $\mathbf{e}_{i}$ at position $i$. Hence, the input to the Transformer block becomes $\mathbf{\hat{e}}_i= \mathbf{e}_{i} + \mathbf{g}_{i}$. By doing so, we differentiate the channel at each subcarrier and assign position dependent attention.
	\subsubsection{Location Identification}
\begin{figure}[!t]
		\centering
		\begin{minipage}{.47\linewidth}
			\begin{subfigure}[t]{.99\linewidth}
				\includegraphics[width=\textwidth]{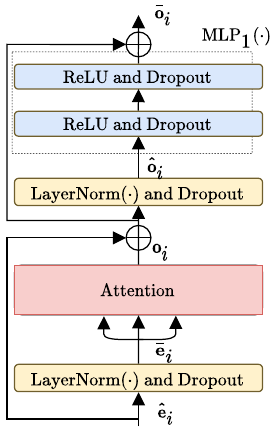}
				\caption{Transformer block.}
				\label{fig:attention_module}
			\end{subfigure}
		\end{minipage}
		\begin{minipage}{.47\linewidth}
			\begin{subfigure}[t]{.99\linewidth}
				\includegraphics[width=\textwidth]{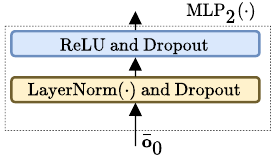}
				\caption{MLP head.}
				\label{fig:mlp_head}
			\end{subfigure} \\
			\begin{subfigure}[t]{.990\linewidth}
				\includegraphics[width=\textwidth]{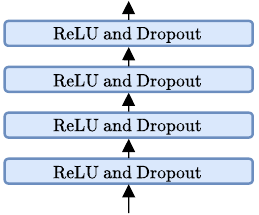}
				\caption{Base DNN.}
				\label{fig:base_dnn}
			\end{subfigure} 
		\end{minipage}
		\caption{DNN model details. a) Transformer block with the attention module for feature learning, b) MLP-head for location estimation, and c) the base-DNN we commonly used in the previous work \cite{9616218}.}
		\label{fig:dnn_model_details}\vspace{-0.1cm}\vspace{-0.1cm}
	\end{figure}

	To add \textit{global} context information on the whole channel, we can prepend to the set of subcarrier embeddings a special symbol $\left[ \verb*|LID| \right] $. This is considered as another learnable vector, $\mathbf{e}_{0} $, whose representation is a compressed characterization of the whole channel from the $r-$th transmitter and it can be used to feed into the multi-layer perceptron (MLP), i.e., the MLP-head detailed in Fig. \ref{fig:mlp_head} for the final features to location mapping. We should note that we investigated averaging over all representations to combine the attended features as the input into the MLP-head, finding $ \left[ \verb*|LID| \right] $ is sufficient but performs worse than the averaging. We report the performance in the Sec. \ref{experiments}. The main reason for using the special vector is future self-supervision and transfer-learning investigation. In the following, we keep the $\left[ \verb*|LID| \right] $. Thus, the set of vectors as input to the transformer block becomes $C = N_{c}' + 1$.
	
	\subsubsection{Attention}
	In case of self-attention \cite{vaswani2017attention}, we consider three input \textit{copies} and project them using the same set of weights, $\mathbf{W}_{q} = \mathbf{W}_{k} = \mathbf{W}_{v}$. To this end; we write the self-attention as \vspace{-0.2cm}
	\begin{equation}
		\begin{aligned}
			\mathbf{o}_{i} = \sum_{j=1}^{C} \frac{\exp \left(\alpha_{i, j}\right)}{\sum_{j^{\prime}=1}^{C} \exp \left(\alpha_{i,j^{\prime}}\right)}\left(\mathbf{\bar{e}}_{j} \mathbf{W}_{v} \right) \\[6pt]
		\end{aligned} \vspace{-0.2cm}
	\end{equation}
	where $\alpha_{i,j}$ is the attention coefficient between the two embeddings at positions $i$ and $j$, \vspace{-0.2cm}
	\begin{equation}
		\begin{aligned}
			\alpha_{i,j} = \frac{1}{\sqrt{D}}\left(\mathbf{\bar{e}}_{i} \mathbf{W}_{q} \right)\left(\mathbf{\bar{e}}_{j} \mathbf{W}_{k} \right)^{T} .
		\end{aligned} \vspace{-0.2cm}
	\end{equation}
	In above, $\mathbf{\bar{e}}_{i} = \mathrm{LayerNorm}(\mathbf{\hat{e}}_{i}; \gamma, \beta) $ where $\gamma$ and $\beta$ are hyperparameters \cite{ba2016layer}.
	\subsubsection{MLP-head}
	At the output of the Transformer block of the proposed model, the representation vector $\mathbf{\bar{o}}_{i}$ is \vspace{-0.1cm}
	\begin{equation}
		\begin{aligned}
			\mathbf{\bar{o}}_{i} = \mathrm{MLP}_{1} \left(  \mathbf{\hat{o}}_{i} \right) + \left( \mathbf{o}_{i} + \mathbf{\hat{e}}_{i} \right) ,
		\end{aligned} \vspace{-0.1cm}
	\end{equation}
	where $\mathbf{\hat{o}}_{i} = \mathrm{LayerNorm} \left( \mathbf{o}_{i} + \mathbf{\hat{e}}_{i}; \gamma, \beta \right)$.
	
	As we discussed, the input to the MLP head can be $\mathbf{\bar{o}}_0$ or an averaged representation over $N_{c}'$ representation vectors. In the case of $\mathbf{\bar{o}}_{0}$, then $\widehat{\mathbf{u}}_{r} = \mathrm{MLP_{2}}(\mathbf{\bar{o}}_{0}) \mathbf{W}_{2} $, where $\mathbf{W}_{2} \in \mathbb{R}^{D\times D'}$ is the weight matrix of the output linear layer, and $D'$ is the number of output units representing the position coordinates.
	
	\section{Simulations and Results}\label{experiments}
	In this section, we evaluate and compare the performance of the proposed approach w.r.t. various factors. In the results, we have labeled this approach as WiT, i.e., Wireless Transformer. Moreover, we discuss a few other aspects encountered during this work. Finally, we conclude this work.\vspace{0.0cm}
	\subsection{Scenarios and Datasets}\label{sec:simulation_parameters}
\begin{figure*}[t]%
		\centering \vspace{0.4584cm}
		\begin{subfigure}{.69\columnwidth}
			\includegraphics[width=1.0\linewidth]{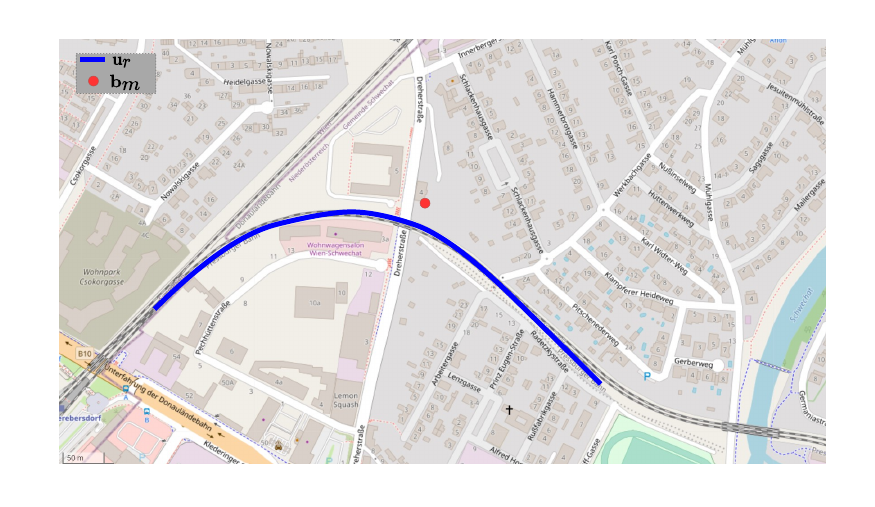}%
			\caption{S-scenario.}%
			\label{fig:s_scenario_osm}%
		\end{subfigure}\hfill%
		\begin{subfigure}{.66\columnwidth}
			\includegraphics[width=\columnwidth]{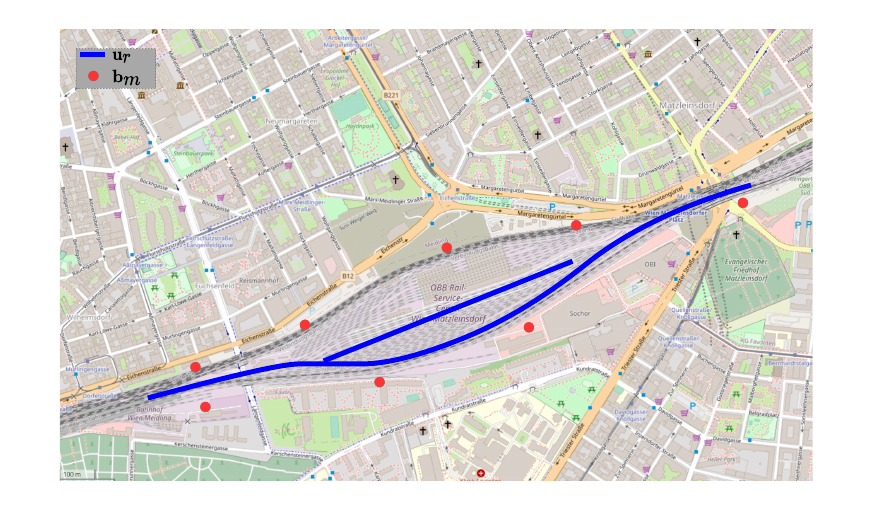}%
			\caption{HB-scenario.}%
			\label{fig:hb_scenario_osm}%
		\end{subfigure}\hfill%
		\begin{subfigure}{.60\columnwidth}
			\includegraphics[width=\columnwidth]{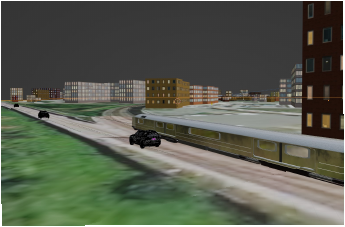}%
			\caption{An example model from S-scenario.}%
			\label{fig:s_scenario_blender}%
		\end{subfigure}%
		\caption{Considered railway trajectories are a) the Schwechat area (S-scenario) and near b) the Vienna central station (HB-scenario). Example of model used for ray-tracing is shown in c). Train moves parallel to the trajectory, and other objects change their position for every $t$ too.}
		\label{fig:OSM_and_Blend}\vspace{-0.5cm}
	\end{figure*}

	\vspace{0.1cm}
	To obtain all the multi-path related parameters for the modeled scenarios, we make use of the available shooting and bouncing ray (SBR) approach with low-angular separation \cite{yun2015ray} in the ray-tracing tool from Matlab \cite{MATLAB2021}. By using the ray-tracer, we are able to simulate the temporal aspects of the scenarios under consideration by simply running $T = 200$ realizations with altered input geometries, changing the position of the considered moving objects, and varying over different material properties as explained in Sec. \ref{systemModel}. The initially imported scenario is from the OpenStreetMap \cite{OpenStreetMap}, and then the 3D tool \cite{blender} is used for modeling the moving objects and changing environment. In this work, we consider two scenarios, as shown in Fig. \ref{fig:OSM_and_Blend}. We assume a single-BS for the first scenario, S-scenario, $M = 1$ and $R = 360$. For the second scenario, HB-scenario, we consider a DAS with $M = 8$ and $R = 406$. In both cases $N_{r} = 64$, $f_{c} = 3.5 \mathrm{GHz}$, $L = 4$, and $B = 20 \mathrm{MHz}$. We consider every $16-$th subcarrier as active, where $N_{c} = 512$, $N_{c}' = 32$ and $N_{c} \equiv N_{c}' \quad(\bmod \; 16)$. The receivers are at a height of $20\mathrm{m}$. Since $u_{r,3} = 1.5\mathrm{m} \; \forall \; r\in \mathcal{R}$, we only consider $D' = 2$ during the training. We consider the default relative permittivity values $\epsilon_{\kappa}$ for $\kappa \in \{\text{concrete, brick, metal, wood}\}$ \cite{series2015effects} and add the atmospheric attenuation in the event of rain with $\mathbb{P}  \left( \boldsymbol{\mathcal{R}} \right) = 0.3$. The obtained sample size is $RT$. However, if the received power is less than $-130 \mathrm{dBm}$, then we discard such measurement at time $t$ from the ray-tracer. Thus, the dataset has a total of $69\,212$ and $81\,200$ samples for S- and HB-scenario, respectively. It is worth noting that the proposed network appears not to saturate within $1800$ epochs, in contrast to the base-DNN \cite{9616218}. However, we limited the training range due to time constraints on limited available resources.\vspace{-0.1cm}
	\subsection{Training Details}\label{sec:training_details}
	\vspace{0.1cm}
	As shown in Fig. \ref{fig:dnn_model_details}, we adopt $\mathrm{ReLU}$ for the intermediate non-linear operations in both $\mathrm{MLP}_{1}(\cdot)$ and $\mathrm{MLP}_{2}(\cdot)$. The proposed network is trained for $1800$ epochs with a batch size of 512. We use Adam solver with weight decay \cite{loshchilov2017decoupled}, and the initial learning rate is set to $3\cdot10^{-4}$. Each layer has $D = 650$ units followed by a dropout rate of $0.1$. For the $\mathrm{LayerNorm(\cdot)}$, the additive factor $\gamma = 1$ and the multiplicative parameter $\beta = 0.0001$. The base-DNN, which we used in \cite{9616218} as a backbone of the then proposed model, consists of four layers, each followed by a dropout with a dropout rate of $0.2$ as detailed in Fig. \ref{fig:base_dnn}. The hidden dimensionality is kept the same, $D$. Early stopping is applied for training the base-DNN if the validation loss does not improve for 80 consecutive epochs. The dataset is split into $0.75$ and $0.25$ for training and holding out validation and testing, respectively. During the training, the location coordinate values, i.e., $u_{r,i}$, are scaled within $[0,1]$. The estimates are scaled back to evaluate the performance. Performance is reported in terms of mean absolute error, $\mathrm{MAE}$, and the $95-$th percentile, \vspace{-0.1cm}
	\begin{equation}
		\begin{aligned}
			\mathrm{MAE} 
			&= \frac{ \sum_{r'=1}^{R_{\mathrm{test}}} \|\widehat{\mathbf{u}}_{r'} - \mathbf{u}_{r'} \|}{R_{\text{test}}}\;.
		\end{aligned} \vspace{-0.1cm}
	\end{equation}
	
	\subsection{Localization Accuracy}
	\vspace{0.1cm}
	Next, we investigate a few aspects that impact localization performance.
	\subsubsection{Static Scenario}
\begin{figure}[!h] 
		\centering 
		\subfloat[S-scenario \label{fig:ecdf_s_scenario}]{%
			\includegraphics[width=0.50\linewidth]{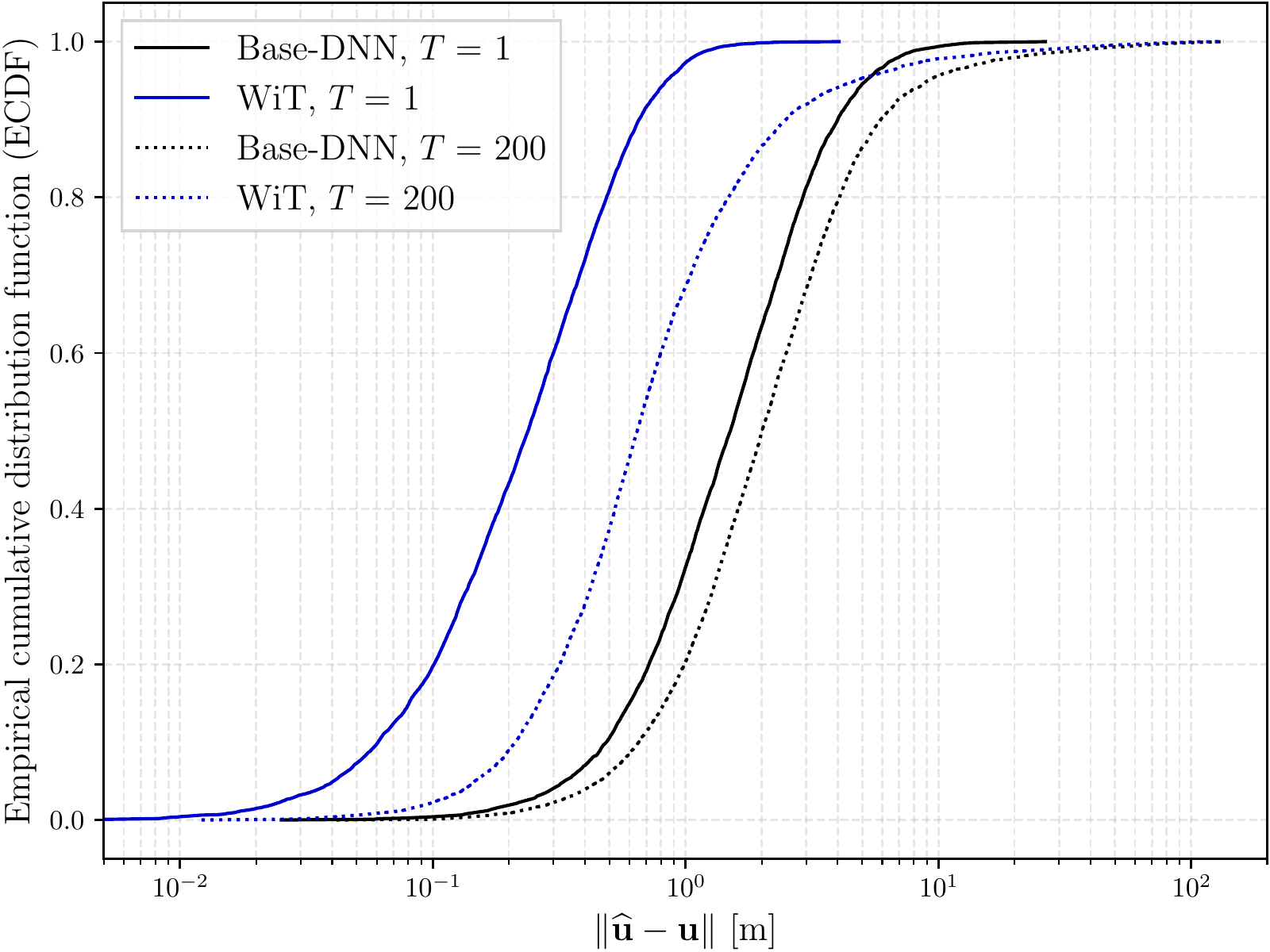}}
		\hfill
		\subfloat[HB-scenario and DAS \label{fig:ecdf_hb_scenario}]{%
			\includegraphics[width=0.50\linewidth]{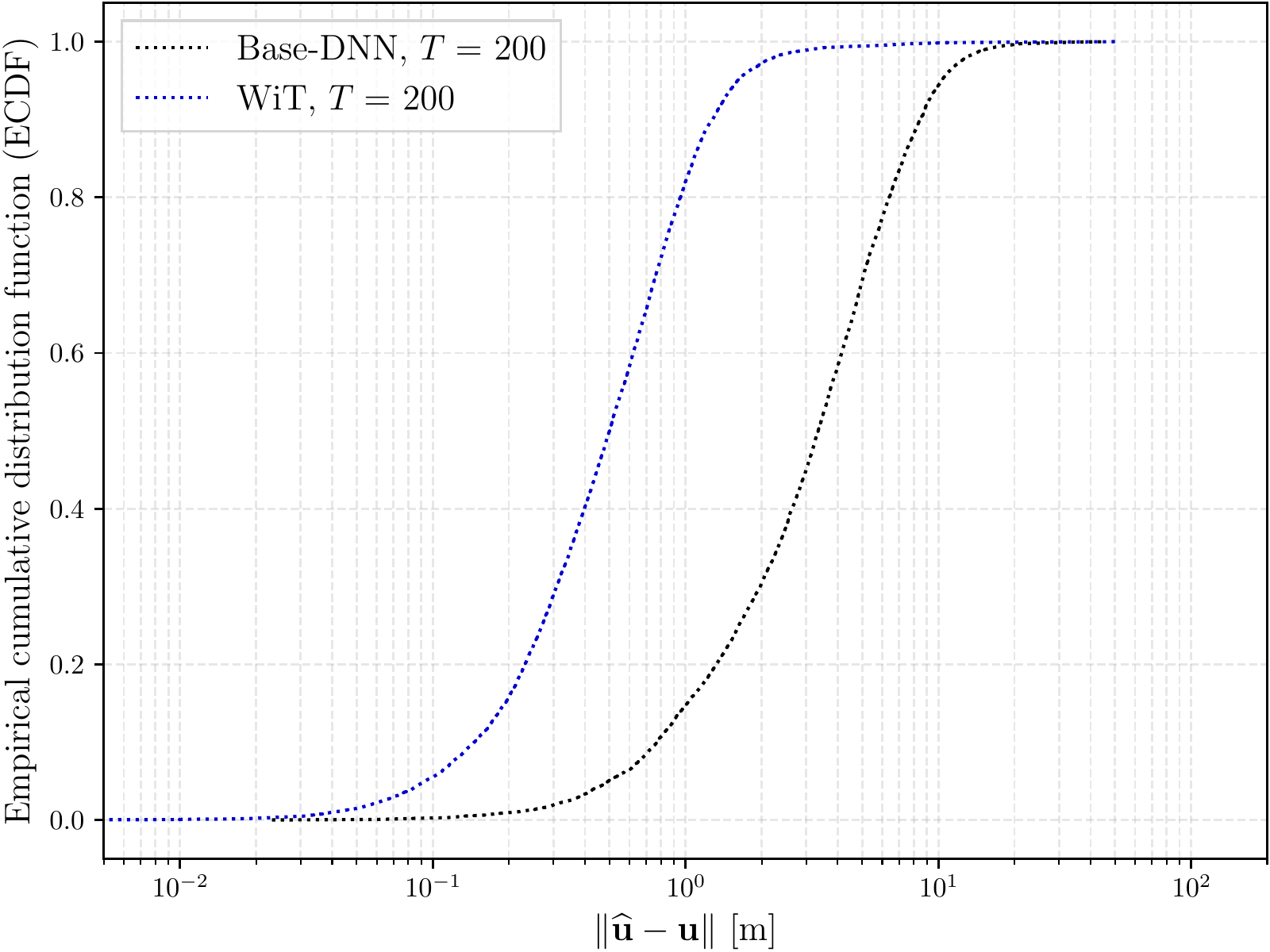}}
		\\
		\hfil
		\subfloat[S-scneario: Actual vs estimated\label{fig:ecdf_s_scenario2}]{%
			\includegraphics[width=0.50\linewidth]{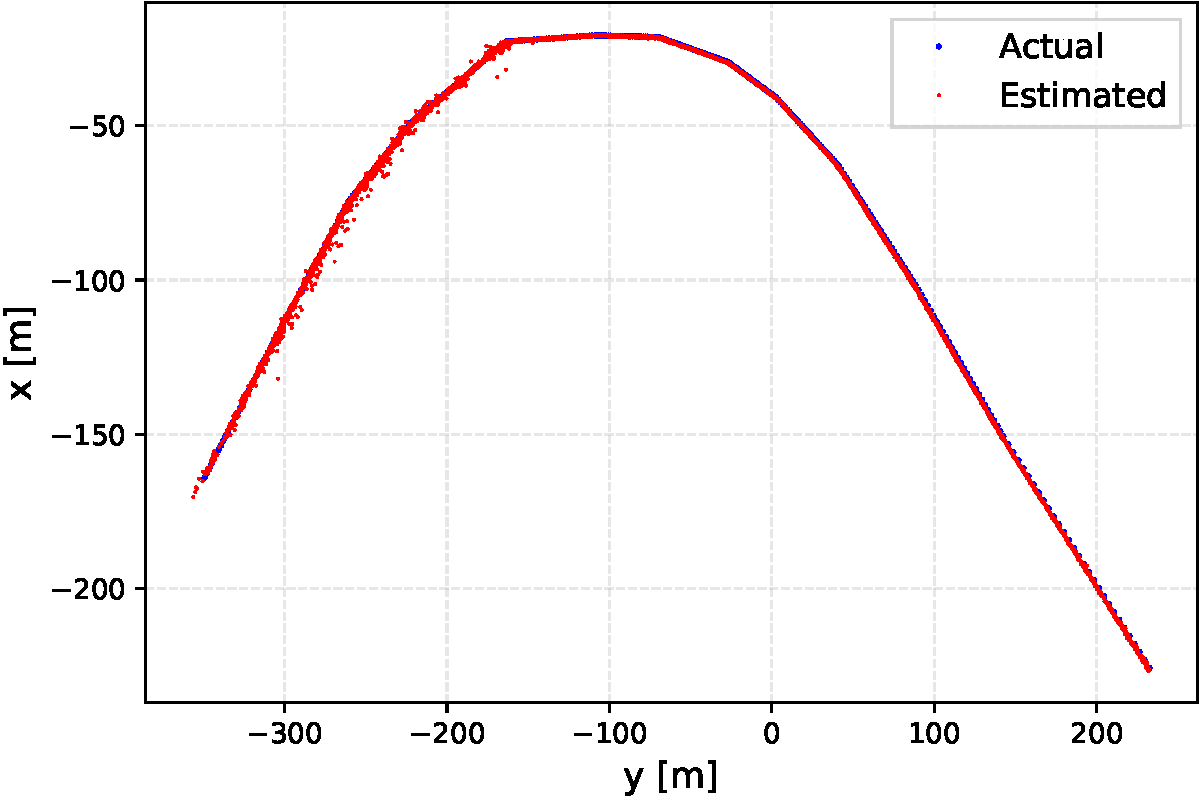}}
		\hfill
		\subfloat[HB-scneario: Actual vs estimated \label{fig:ecdf_hb_scenario2}]{%
			\includegraphics[width=0.50\linewidth]{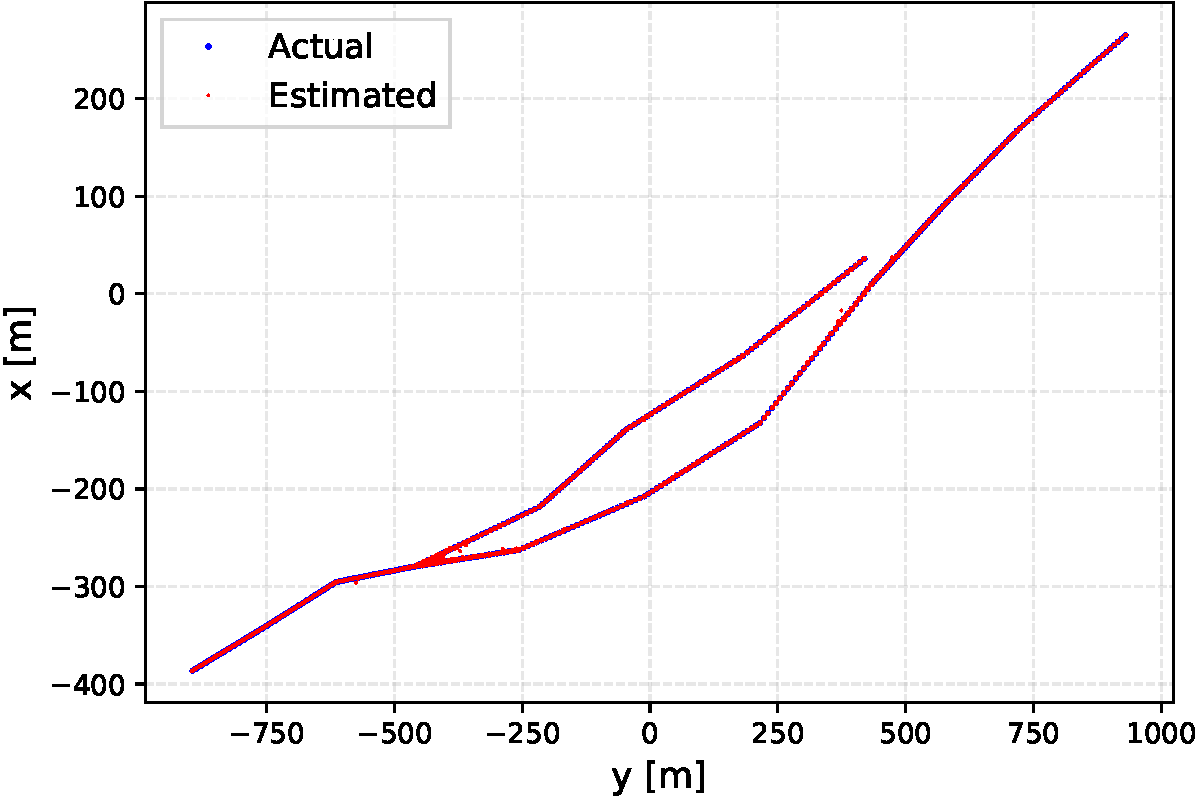}}
		\vspace{0.0cm}\caption{Localization error in (a) S-scenario for $T = 1$ and for $T=200$ and (b) HB-scenario for $T = 200$ time snapshots. Proposed learned features outperform the base DNN across all the locations. In c) and d) actual versus estimates in the trajectory for $T = 200$ are shown. }
		\label{fig:ecdf_error_scenarios}\vspace{-0.5cm}
	\end{figure}

	First, we investigate the impact of the learned features in the S-scenario for the static environment, $T=1$. To have a sufficient amount of training samples, the inter distance between any two locations $\| \mathbf{u}_{i} - \mathbf{u}_{j} \| $ is much smaller than that of a dynamic scenarios. Thus, the dataset for this case consists of $72\,000$ channel and location pairs. The attention based features consistently perform better compared to the raw CSI and a base-DNN. The accuracy is improved by more than $50\%$. The localization performance, comparison to the base-DNN, and comparison to the actual test locations in $\mathbb{R}^{2}$ is depicted in Figs. \ref{fig:ecdf_s_scenario}, \ref{fig:ecdf_hb_scenario},  \ref{fig:ecdf_s_scenario2} and \ref{fig:ecdf_hb_scenario2}.
	
	\subsubsection{Mobility Scenario}
	Similarly, for the dynamic scenarios and $T = 200$, the proposed features, learned by the attention mechanism, are much more robust than the raw CSI and the base-DNN, reducing the localization error by a significant margin.
	
	\subsubsection{Distributed Antennas}
	As we mentioned earlier, for the HB-scenario we consider $N_r$ distributed antennas among $M = 8$ infrastructure nodes. From Fig. \ref{fig:ecdf_hb_scenario}, we can observe that the proposed approach provides significantly better performance.
	
	\subsubsection{Impact of Features Averaging}
	Table \ref{table: results_table} shows the summarized results and the performance gap when comparing the averaging over features and the case of using the unique representation vector for each set of subcarriers. Again, the performance difference is evident in both DAS and single-BS scenarios. Yet, this compressed representation version can outperform the base-DNN.
	\vspace{0.0cm}
	
	\begin{table}[!t]
		\captionsetup{size=small,
			skip=5pt, position = bottom}
		\caption{Summarized Results}
		\scalebox{0.95}{%
			\begin{tabular}{lcccccccc}
				\hline & \multicolumn{2}{c}{ S } & & \multicolumn{2}{c}{ S } & & \multicolumn{2}{c}{ HB} \\
				& \multicolumn{2}{c}{ $T=1$ } & & \multicolumn{2}{c}{ $T=200$ } & & \multicolumn{2}{c}{ $T=200$ } \\
				\cline { 2 - 3 } \cline { 5 - 6 } \cline { 8 - 9 } \textbf{Method} & MAE & $95-$th & & MAE & $95-$th & & MAE & $95-$th \\
				\hline Base-DNN & $1.98$ & $5.16$ & & $3.59$ & $8.83$ & & $4.13$ & $10.01$ \\
				WiT $\mathrm{[LID]}$ & $0.74$ & $1.88$ & & $2.36$ & $6.54$ & & $1.18$ & $2.83$ \\
				\hline WiT (avg.) & $\mathbf{0.31}$ & $\mathbf{0.84}$ & & $\mathbf{1.70}$ & $\mathbf{4.70}$  & & $\mathbf{0.68}$ & $\mathbf{1.61}$ \\
				\hline
		\end{tabular}}
		\label{table: results_table} \vspace{-0.4cm}
	\end{table}
	\subsection*{Discussion}\label{Discussion}
	\vspace{0.0cm}
	A naive application of the attention mechanism would involve inputting every channel coefficient into the network, such that each real-valued channel coefficient \textit{attends} to every other. With the increased number of antenna elements and subcarriers in massive MIMO, this would not scale to realistic future systems. Still, one of the critical challenges of utilizing the attention mechanism on a more extensive set of subcarriers is its efficiency due to the computation and memory complexity. Furthermore, we have noticed that the applied residual connections play a crucial role in retaining the position information on the subcarriers representation after the attention module. Without the residual connection, the information about the original structure of the channel is lost. Removing the residual connections might lead to the loss of such information after the attention module. Moreover, with randomly initialized parameters for self-attention vectors, the position has no relation to its original input. \vspace{-0.2cm}

	\section{Conclusion}\label{conclusion}
	\vspace{-0.1cm}
	We presented an end-to-end and DNN-based localization method with robust feature learning. We proposed to input each subcarrier into the network, and using the attention mechanism we were able to better capture the dependence in the CSI over the subcarriers, providing superior localization performance compared to the base-DNN with raw CSI. We investigated dynamic scenarios where the scattering events over $T$ time snapshopts cause time variations in the channel. We showed that the proposed method is able to cope with imperfect channel estimates. In this work, we also modeled two ray-tracing based scenarios over railway tracks. Finally, we showed that the proposed method excels by even a greater margin when a distributed antenna system is considered.
	\vspace{-0.1cm}
	\section*{Acknowledgment}
	\vspace{0.00cm}
	This work has been funded by \"OBB Infrastruktur AG. The financial support by the Austrian Federal Ministry for Digital and Economic Affairs, the National Foundation for Research, Technology and Development and the Christian Doppler Research Association is gratefully acknowledged.
	\vspace{0.1cm}
	
	\bibliographystyle{IEEEtran}
	\bibliography{References}
\end{document}